# A $55 Shock Tube for Simulated Blast Waves

Elijah Courtney, Amy Courtney, Michael Courtney

BTG Research, 9574 Simon LeBleu Road, Lake Charles, LA, 70607
michael_courtney@alum.mit.edu

**Abstract**
Shock tubes are commonly employed to test candidate armor materials, validate numerical models, and conduct simulated blast experiments in animal models. As DoD interests desire to field wearable sensors as blast dosimeters, shock tubes may also serve for calibration and testing of these devices. The high blast pressures needed for experimental testing of candidate armors are unnecessary to test these sensors. An inexpensive, efficient, and easily available way of testing these pressure sensors is desirable. It is known that releasing compressed gas suddenly can create a repeatable shock front, and the pressures can be finely tuned by changing the pressure to which the gas is compressed. A Crosman 0.177 caliber air pistol was used (without loading any pellets) to compress and release air in one end of a 24" long 3/4" diameter standard pipe nipple to simulate a blast wave at the other end of the tube. A variable number of pumps were used to vary the peak blast pressure. As expected, the trials where 10 pumps were used to compress the air resulted in the largest average peak pressure of 101.99 kPa (± 2.63 kPa). The design with 7 pumps had the second biggest peak pressure, with an average peak pressure of 89.11 kPa (±1.77 kPa). The design with 5 pumps had the third largest peak pressure, with an average peak pressure of 78.80 kPa (±1.74 kPa). 3 pumps produced an average peak pressure of 61.37 kPa (±2.20 kPa). 2 pumps produced an average peak pressure of 48.11 kPa (±1.57 kPa). The design with just 1 pump had the smallest peak pressure and produced an average peak pressure of 30.13 kPa (±0.79 kPa). This inexpensive shock tube design had a shot-to-shot cycle time of between 30 and 45 seconds.

Key words: shock tube, blast, blast injury, armor, traumatic brain injury, blast dosimeter

**Introduction**
With the increase in traumatic brain injury (TBI) over the last two decades, the number of laboratory experiments investigating blast waves has increased dramatically. However, the prohibitive cost and difficulty of setting up and using shock tubes prevents many institutions and individuals from investigating blast wave effects. In recent years, there has been substantial progress made with regard to making shock tubes less expensive and easier to obtain and use. The table-top shock tube (Courtney and Courtney, 2010) utilized a small-scale shock tube capable of creating pressures up to 3.8 MPa. A combustion-driven shock tube, fueled by deflagrating oxy-acetylene to produce a simulated blast wave, may be used to test candidate armor materials, validate numerical models, and apply simulated blast waves to animal models (Courtney et al., 2012). A combustion-driven shock tube was also used to produce a blast wave via detonation of oxy-acetylene (Courtney et al., 2014), generating pressures of more than 5 MPa. Extensive numerical modeling of a blast-driven shock tube has been done (Alley, 2009), along with numerical modeling of compression-driven shock tube designs (Chandra et al., 2012).

Both blast driven and compression driven shock tubes are commonly used for testing armor, blast experiments in animal models, and for validating numerical models of blast wave effects. However, though effective, these devices tend to be expensive, with many administrative and legal obstacles to implementing certain shock tube designs in university and other research environments. In addition, these devices produce pressures in excess of those necessary for developing, testing, and calibrating blast dosimeters. Pressures that are between 0 and 350 kPa are needed for blast dosimeter testing (Kahn, 2007). Many shock tubes are designed to apply higher peak blast pressures.

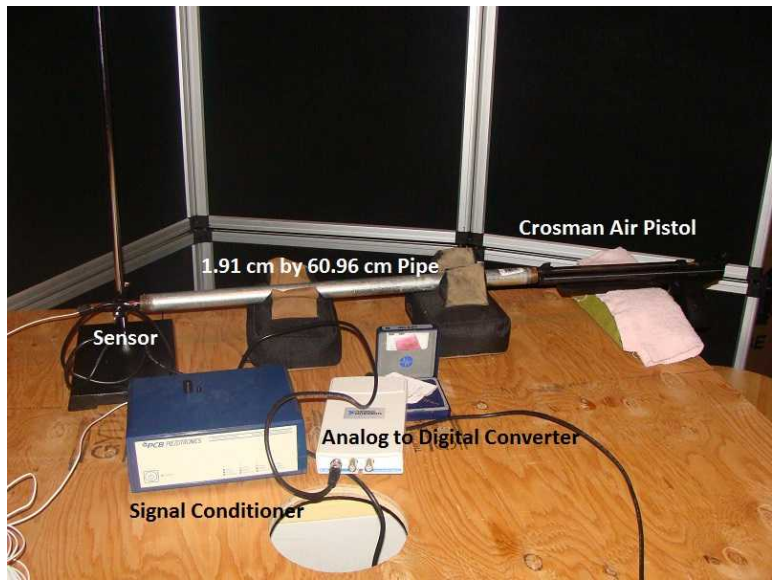

*Figure 1: Test setup for experiments*.

In this experiment, a small-scale compression-driven shock tube was created to consistently produce a range of peak blast pressures, ranging from about 30 kPa to about 100 kPa. These pressures are adequate for testing and calibrating piezoelectric blast dosimeters. The shock tube cost less than $60, has a shot-to-shot cost of nearly nothing, and can produce pressure waves every 35-40 seconds. Previously, the least expensive and fastest shock tubes cost more than $1000 (Courtney et al., 2014), had a shot-to-shot cost of about $10, with a shot-to-shot cycle time of about 10 minutes. Other explosive driven and compressed gas driven shock tubes can have much higher costs, overhead, and required approvals and may only produce a limited number of trials per day. Previous shock tube designs are also very loud and require some combination of insulation and isolation to prevent disturbing workers throughout many buildings. The shock tube in this experiment requires few safety precautions other than the common sense associated with an air pistol and is no louder than the ordinary firing of an air pistol.

**Materials and Method**

In this design, the driving section of the shock tube was a Crosman Pumpmaster Classic 0.177 caliber air pistol. The driven section was a 1.91 cm wide by 60.96 cm long standard piece of steel pipe. After the air had been compressed using the lever mechanism, the driving section rested on a sandbag, and was placed flush against one open end of the metal pipe. As the compressed air was released via the trigger, the corresponding air jet achieved supersonic speeds in the barrel, producing a shock front which evolved into a simulated blast wave as it traveled along the pipe.

For this experiment, a high-speed piezoelectric pressure sensor (PCB Piezotronics 102B15) was secured by a ring stand and placed flush against the opposite end of the pipe with its face perpendicular to the direction of travel of the blast wave. Five trials for each number of pumps were recorded, to better measure the simulated blast wave and quantify shot to shot variations. Pressure data was recorded at a sample rate of 1 MHz via cables which connected the pressure transducer to a signal conditioning unit (PCB 842C), producing a voltage output. This output was then digitized with a National Instruments USB-5132 fast analog to digital converter and stored in a laptop computer. Digitized voltage vs. time data was then converted to pressure vs. time using the calibration certificate provided by the manufacturer of the pressure sensor.

*Table 1: The average peak pressures at the shock tube opening for each number of pumps, along with standard deviations and the standard error of the mean (SEM).*

| Results | | | | | |
|---|---|---|---|---|---|
| **Number of Pumps** | **Average Peak Pressure (kPa)** | **Standard Deviation (kPa)** | **Standard Deviation (%)** | **SEM (kPa)** | **SEM (%)** |
| **1 Pump** | 30.13 | 0.79 | 2.62 | 0.35 | 1.16 |
| **2 Pumps** | 48.11 | 1.57 | 3.26 | 0.70 | 1.45 |
| **3 Pumps** | 61.37 | 2.20 | 3.58 | 0.99 | 1.61 |
| **5 Pumps** | 78.80 | 1.74 | 2.21 | 0.78 | 0.99 |
| **7 Pumps** | 89.11 | 1.77 | 1.99 | 0.79 | 0.89 |
| **10 Pumps** | 101.99 | 2.63 | 2.58 | 1.18 | 1.16 |

**Results**

Table 1 shows the mean peak pressure, standard deviation, and the standard error of the mean for the varying number of pumps for this design. The duration of the pressure wave produced by each design stayed fairly constant at about 0.05 ms. The pressure profile differs slightly from the Friedlander waveform that often characterizes free field blast waves. The simulated blast wave takes several microseconds to reach peak pressure, the decay does not seem to approximate an exponential decay, and the waveform does not reliably reproduce a negative pressure phase. These features may be suboptimal for use in animal models, but are not prohibitive from applications calibrating and validating piezoelectric dosimeters which sample the wave shape and produce output as a digitized voltage vs. time representing the incident blast pressure.

Figure 2 shows pressures measured at the shock tube opening by the pressure sensor with its face perpendicular to the direction of travel of the blast wave at a distance of 0 mm. The shape and magnitude of the blast wave were repeatable for each of the various numbers of pumps. Each graph is not the average of the 5 trials, but is rather the trial with the peak pressure closest to the average for those trials. Earlier work with shock tubes suggests that the temporal profile of the simulated blast wave degrades rapidly at distances beyond one half of a diameter in front of the tube opening. Consequently, it is recommended in this case that the test object be located within 0.9 cm of the shock tube opening.

Figure 3 shows the average peak pressures and the standard error of the mean for each number of pumps. Since it is expected that the peak pressure in a shock wave is proportional to the square root of the energy, the measured peak blast pressure was fit to a function given by a constant times the square root of the number of pumps of the air pistol. The model fit the data well giving, P(x) = 33.22 kPa sqrt(x), where x is the number of pumps and P is the mean peak pressure produced. The $R^2$ of the least squares fit was 0.9842.

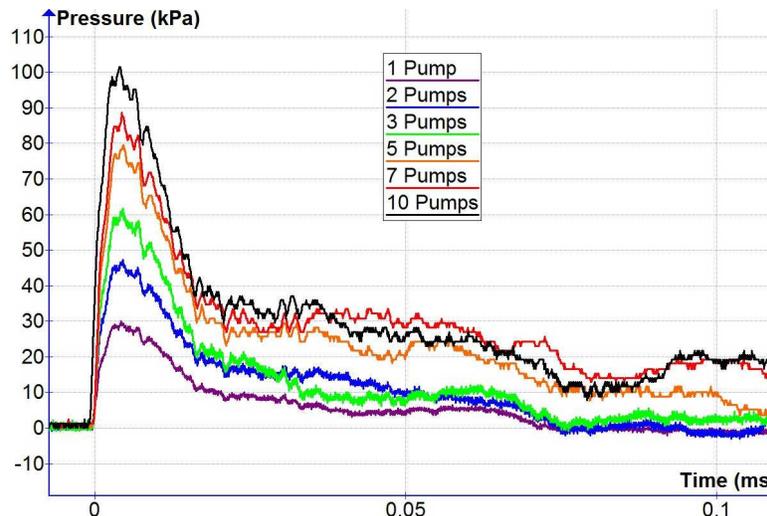

*Figure 2: Blast pressure as a function of time produced by the shock tube for various numbers of pumps.*

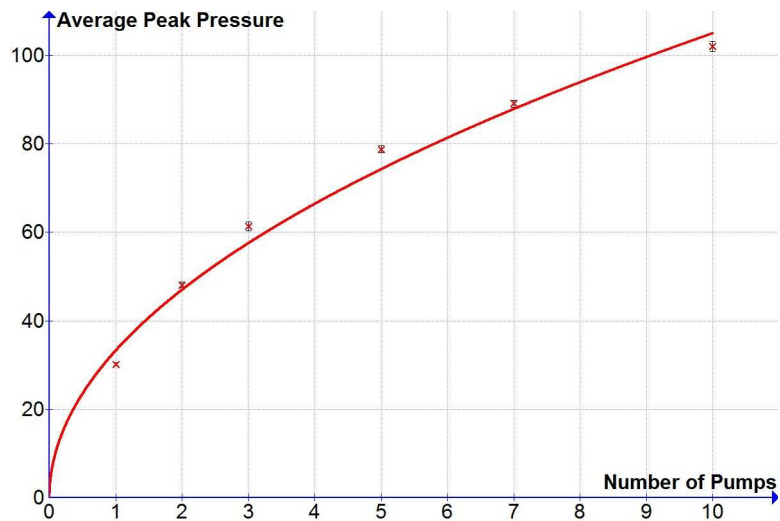

*Figure 3: The average peak pressures for each number of pumps.*

**Discussion**

As described in the introduction, there are several reasons why this design may be more appropriate for some applications than larger shock tubes. Further, given the variety of inexpensive, compression driven projectile launchers which quickly release a measured and consistent amount of gas when the trigger is pressed, it is likely that a suitable shock tube can be quickly developed for an even wider variety of applications. Perusing the aisles at Walmart reveals a variety of pump and compressed $CO_2$ driven .177 and .22 caliber air rifles and pistols, a variety of paintball markers in various calibers, and a variety of Airsoft pistols and rifles. Matching an appropriate projectile launcher (with no projectile) to serve as the driving section of a shock tube with an appropriate length and diameter of pipe to serve as the driven section should be able to produce a wide variety of blast waves for various applications.

Some limitations are that this design does not produce pressures suitable for testing candidate armor materials and the peak pressures are at the lower end of the range commonly used for validating numerical models and conducting traumatic brain injury experiments in animal models. However, the authors feel that because these functions are easily filled by other, larger shock tubes, this does not significantly impact the usefulness and efficiency of the design. Another limitation of this design (and many compression driven shock tubes) is the "jet effect" whereby the compressed gas flows past the object under test and may transfer momentum that is atypical of genuine blast waves. Our earlier shock tube designs (Courtney et al., 2012; Courtney et al., 2014) based on detonation or deflagration mitigated this effect because the combustion products occupied less volume at standard temperature and pressure than the reactants; however, in this design, the venting gases will transfer more momentum to the object under test than more realistic blast waves. This extra momentum transfer is tolerable in applications calibrating and/or validating wearable dosimeters, but may introduce unrealistic injury mechanisms if used in animal models.

**Acknowledgments**

Andrew Barraford of the University of Massachusetts, Amherst contacted us in late 2014 about testing their emerging wearable dosimeter designs with one of our existing shock tube designs. We realized that our existing designs were too expensive, too loud, and too slow for the calibration and validation of wearable dosimeters in a production environment. Recognizing that most compression driven shock tubes operate by the sudden release of compressed gas at one end of a tube, we thought an unloaded air pistol might fulfill this role nicely. Campus regulations prevented our invention from being adopted in their laboratory, but Andrew Barraford and his colleagues recognized the important principles and soon developed an appropriate shock tube for their needs. We are grateful for their inspiration in sharing their designs and needs and results with us in a way that motivated the current invention. This shock tube design continues the theme of realizing laboratory shock tubes as an existing projectile launcher without the projectile (and with other potential modifications.) Our first shock tube invention was envisioned as a potato cannon without a potato but then modified to use steel pipe and oxy-acetylene after watching a golf ball launcher video by Ian Morris (https://www.youtube.com/watch?v=hafdriLJY7w ).